\documentclass[sn-basic]{sn-jnl}

\usepackage{graphicx}%
\usepackage{amsmath,amssymb,amsfonts}%
\usepackage{amsthm}%
\usepackage{mathrsfs}%
\usepackage[title]{appendix}%
\usepackage{xcolor}%
\usepackage{textcomp}%
\usepackage{manyfoot}%
\usepackage{booktabs}%
\usepackage{algorithm}%
\usepackage{algorithmicx}%
\usepackage{algpseudocode}%
\usepackage{listings}%

\usepackage{hyperref}   
\hypersetup{
  colorlinks=true,
  citecolor=blue,
  linkcolor=blue,
 urlcolor=blue}
\usepackage{booktabs}\let\cline\cmidrule
\usepackage{multirow}
\usepackage{amsfonts}
\usepackage{amsmath}

\theoremstyle{thmstyleone}%
\theoremstyle{thmstyletwo}%

\theoremstyle{thmstylethree}%

\begin{document}

\title[Article Title]{Bayesian and Convolutional Networks for Hierarchical Morphological Classification of Galaxies}


\author*[1]{\fnm{Jonathan} \sur{Serrano-Pérez}}\email{js.perez@inaoep.com}

\author[2]{\fnm{Raquel} \sur{Díaz Hernández}}\email{raqueld@inaoep.mx}

\author[1]{\fnm{L. Enrique} \sur{Sucar}}\email{esucar@inaoep.mx}

\affil*[1]{\orgdiv{Computer Science Dept.}, \orgname{Instituto Nacional de Astrofísica, Óptica y Electrónica}, \orgaddress{\street{Luis Enrique Erro No. 1}, \city{San Andrés Cholula}, \postcode{72840}, \state{Puebla}, \country{Mexico}}}

\affil[2]{\orgdiv{Optics Department}, \orgname{Instituto Nacional de Astrofísica, Óptica y Electrónica}, \orgaddress{\street{Luis Enrique Erro No. 1}, \city{San Andrés Cholula}, \postcode{72840}, \state{Puebla}, \country{Mexico}}}


\abstract{
This work is focused on the morphological classification of galaxies following the Hubble sequence in which the different classes are arranged in a hierarchy. The proposed method, BCNN, is composed of two main modules. First, a convolutional neural network (CNN) is trained with images of the different classes of galaxies (image augmentation is carried out to \textit{balance} some classes); the CNN outputs the probability for each class of the hierarchy, and its outputs/predictions feed the second module. 
The second module consists of a Bayesian network that represents the hierarchy and helps to improve the prediction accuracy by combining the predictions of the first phase while maintaining the {\em hierarchical constraint} (in a hierarchy, an instance associated with a node must be associated to all its ancestors), through probabilistic inference over the Bayesian network so that a consistent prediction is obtained.
Different images from the Hubble telescope have been collected and labeled by experts, which are used to perform the experiments. The results show that BCNN performed better than several CNNs in multiple evaluation measures, reaching the next scores: 67\% in exact match, 78\% in accuracy, and 83\% in hierarchical F-measure.
}




\maketitle

\section{Introduction}

Galaxy morphological classification is essential for understanding galaxies evolution and studying stellar populations and their physical properties. For the classification, large-scale data sets are needed. Galaxy classification continues to face several difficulties and challenges despite recent advances. Some current issues and challenges include subjectivity that can lead to inconsistencies and variations in classification results, mainly due to subjective decisions made by experts based on visual inspection. The ambiguity in the diverse and complex morphology that some galaxies may present makes it difficult to define clear boundaries between the classes, allowing overlap between them, which hinders accurate classification. 
On the other hand, data set biases can affect the generalization and performance of models used for classification. Well-selected and balanced data sets for training and evaluation of the models are crucial for accurate classification. 

In our experience, image quality is one of the challenges we face in classifying galaxy morphology. As the galaxy's images are taken using telescopes, the image quality depends on observation techniques, sky section selection, and exposition time. When galaxies are presented face-on and are well-centered, classification is more straightforward. However, it is more difficult to analyze their structure when they are edge-on, tilted, or in other positions. Therefore, it is necessary to develop identification and classification techniques that are invariant concerning the position and size of the galaxies. Also, galaxy classification models trained on specific data sets may have difficulty generalizing well to new and unknown data sets.

The challenge is to build and train computational models that can effectively capture the inherent characteristics of galaxies across different surveys, instruments, or observing conditions. As data sets grow in size and complexity, accuracy and computational efficiency become significant challenges. 

An example of a large dataset to classify is \cite{Cheng-2022-Lessons}, where a comparison of two comprehensive galactic morphology catalogs created with Convolutional Neural Networks (CNNs) using Dark Energy Survey (DES) data is performed. Despite methodological differences, the two catalogs agree well up to certain magnitudes, demonstrating the reliability of CNN predictions for faint samples. The combined use of both catalogs allows the identification of unusual galaxies, providing valuable insights into galactic morphology and evolution. In a prior publication by the same author, the challenges in galaxy morphological classification are discussed, emphasizing the impact of decreasing apparent brightness and size on the accurate assessment of morphological details. The study also addresses the necessary bias corrections to address these challenges. A comparison is made between classifications obtained through CNNs and visual assessments, revealing discrepancies and underscoring the need for corrections in the galaxy sample \citep{Cheng-2021-GMCC}.

On the other hand, using deep learning algorithms, namely CNNs, another study utilized two classification schemes: T-type, related to the Hubble sequence, and the Galaxy Zoo 2 (GZ2) classification. The research generated the most extensive and precise morphological catalog to date by integrating precise visual classifications \citep{Dominguez_Sanchez-2018-IGM-SDSS-DL}. The performance of these models outperformed previous models trained with support vector machines. T-type classifications showed lower deviation and dispersion than earlier models used for this purpose.

Advances in the field involve the development of hybrid models that leverage data-driven approaches and physical modeling to improve classification accuracy. Furthermore, endeavors to assemble inclusive and comprehensive datasets, mitigate biases, and enhance deep learning models, can be crucial in solving the existing challenges in galaxy classification \citep{huertas-2023-DawesReview, Goan-2020-CSABDS}.

A particular case of solution proposes to use CNNs and Bayesian Neural Networks (BNNs) \citep{Goan-2020-CSABDS}. Thus, using these techniques in morphological galaxy classification, the networks will estimate the properties of the galaxies, which, in turn, will allow for a better morphological classification. 

The great utility of deep learning in astronomy is straightforward, specifically for galaxy classification. However, previous work does not take advantage of the hierarchy, which could help to improve accuracy. We propose Bayesian and Convolutional Neural Networks (BCNN) for morphological galaxy classification in the present work. This method combines a CNN trained with different classes of galaxies and a Bayesian network that represents the hierarchy of each type analyzed by the CNN.

The Bayesian network helps to improve the prediction accuracy by combining the predictions of the CNN while maintaining the {\em hierarchical constraint} (in a hierarchy, an instance associated with a node must be associated with all its ancestors) through probabilistic inference over the Bayesian network so that a consistent prediction is obtained.
Also, we solve the class imbalance problem using data augmentation through geometric transformations.

\subsection{Related Work} \label{s:RW}

\begin{figure}[h!] 
	\centering
    \includegraphics[width=.3\columnwidth]{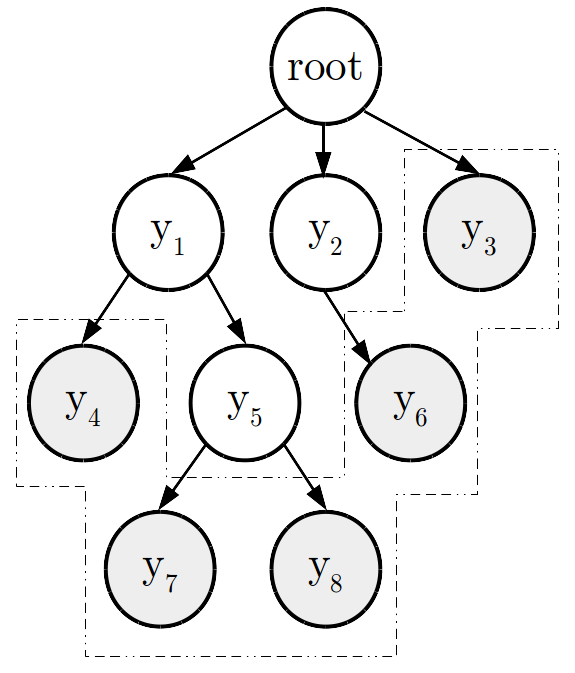}
    \caption{Hierarchical classification considers, in training and prediction phases, all nodes and their relations in the hierarchy $\{y_{1},...,y_{8}\}$. \textit{Flat} classification focus its training and prediction phases only on the leaves nodes (shaded in gray) $\{y_{3},y_{4},y_{6},y_{7},y_{8}\}$ while ignoring the rest of the nodes. }
  \label{f:hierVSflat}
\end{figure}

In hierarchical classification there is a hierarchy where the labels (nodes) are arranged, an example of hierarchy is shown in Fig. \ref{f:hierVSflat}. Hence, hierarchical classification considers the hierarchy (and the relations among the labels) in its training and predictions phases. On the other hand, \textit{flat} classification focuses its training and predictions phases only on the most specific labels (the leaves of the hierarchy) while ignoring the rest of the hierarchy, as shown in Fig. \ref{f:hierVSflat}.

Few works address the problem of hierarchical classification of galaxies since most of them are focused on \textit{flat} classification, that is, they do not use a hierarchy, or there is no hierarchy to perform hierarchical classification.

\cite{Bazell-2000-FRiMG} carried out an analysis of feature selection for galaxy classification. Seven classes of galaxies are considered in the paper; some are also grouped to reduce the classifier's output; 22 features were selected, and an artificial neural network was trained with back-propagation as the classifier.
Later, \cite{Bazell-2001-EoCfMGC} extended the previous work by including different classifiers, such as naive Bayes, decision trees, and artificial neural networks. Additionally, experiments with bagged ensembles were carried out, showing that an ensemble may help improve a single classifier's performance; the best performance was obtained with neural networks along bagged ensembles, where the classification error was $\sim 12\%$ when only two classes are considered.

\cite{Selim-2017-NMF-GC} proposed a scheme based on the Nonnegative Matrix Factorization algorithm to perform morphological classification of galaxies. Four types of galaxies were considered: elliptical, lenticular, spiral, and irregular. Their method was applied to two datasets, small and large, where they report scores of $\sim 93\%$ and $\sim 92\%$ on accuracy, respectively.

\cite{Diaz-Hernandez-2016-AAtStMGCPUtSRTaDL} addressed the problem of galaxy classification using the sparse representation technique and dictionary learning with \textit{K-SVD} \citep{Aharon-2006-KSVD}. They extracted 20 features from each galaxy related to the galaxies' shape, intensity, and area. They consider eight types of galaxies which are clustered from two to seven classes, where they find that the most challenging group to qualify is the one of spiral galaxies. The highest accuracy was obtained when two classes were considered, $\sim 82\%$ on accuracy, nevertheless, when considered the seven classes the performance was decreased to $\sim 44\%$.

\cite{Khalifa-2018-DeepGalaxyV2} proposed Deep Galaxy V2. The method consists of a convolutional neural network formed by eight layers; four are used to perform feature extraction, while the rest are focused on classification. Results in the paper reach up to $\sim 97\%$ in exact match; nevertheless, they only consider three classes of galaxies: elliptical, spiral, and irregular.

\cite{Marin-2013-AHMfMGC} carried out a study of the classification of galaxies using a hierarchy. They used the same features as \cite{Bazell-2000-FRiMG} plus some geometric moments. Their classification scheme follows two steps; in the first step, they try to identify \textit{stars} and \textit{galaxies}. In the second step, they can classify the previously identified galaxies into a specific class of galaxies. They also show that their hierarchical approach got better results than a flat approach, that is, $\sim 53\%$ and $\sim 39\%$, respectively, in exact match.

This paper proposes a novel approach, \textit{Bayesian and convolutional neural networks}, consisting of a Bayesian network fed by a CNN. Unlike most related works, there is no feature extraction process to train a classifier later; instead, a pre-trained CNN is used to extract image features automatically. In general, the features learned automatically by a CNN tend to give better results than manually selected features; additionally, it gives us the advantage that the training can only be focused on the last layer of the CNN. Furthermore, the hierarchy is modeled as a Bayesian network, a novel way to represent the hierarchy and enforce the hierarchical constraint that related methods do not consider.
On the other hand, results of the related work can not be compared directly against our results due to several reasons: the datasets are not the same, most works make use of different variants of the datasets, and they do not usually share them; the labels/classes are not the same, some works focus their research on the most general labels (spiral, irregular, elliptical, lenticular) while in this work there are 10 different labels (arranged in a hierarchy).

The rest of the document is organized as follows. Section \ref{s:fundamentals} summarizes the fundamentals of hierarchical classification, the standard evaluation measures, CNNs, and Bayesian networks; and presents the proposed classifier. Section \ref{s:EandR} introduces the galaxies dataset and shows the experiments and results. Finally, in section \ref{s:CandFW}, conclusions and some ideas for future work are given.

\section{Methodology} \label{s:fundamentals}
\subsection{Fundamentals} 
In hierarchical classification, there are several labels to which an instance can be associated. However, the labels are arranged in a predefined structure, a hierarchy, which commonly is a tree but, in its general form, is a directed acyclic graph. The hierarchy, $H$, can be denoted with graph notation: $H=(L,E)$, where $L$ is the set of labels or nodes and  $E$ is the set of edges that links the nodes. Furthermore, the labels associated with an instance must comply with the \textit{hierarchical constraint}, which states if an instance is related to a node, the instance must be associated with all the ancestors of that node given by the hierarchy. Therefore, a set of labels that do comply with the hierarchical constraint is called \textit{consistent path} or \textit{consistent prediction}. 

Even though there are several hierarchical problems \citep{Silla2011}, in this work, the problem of interest is described as a hierarchy of tree type, where the instances are associated with a single path of labels (SPL) that always reach a leaf node, that is, full depth (FD); $<$Tree, SPL, FD$>$  following  \cite{Silla2011} notation. 

Therefore, the problem of hierarchical classification consists in assigning to a particular object described by $d$ attributes, a subset of labels that comply with the hierarchical constraint: $f_{HC}=\mathbb{R}^{d} \rightarrow  \{0,1\}^{\mid L \mid }$.\\

One key element in this work is the use of Bayesian networks to represent the hierarchy. Bayesian networks are directed graphical models representing the joint distributions of a set of random variables \citep{Sucar-2021-PGMbook}. In a Bayesian network, nodes represent the random variables, while arcs represent direct dependencies among the nodes. In other words, the graph structure encodes a set of independent relations among variables.  

Let the Bayesian network $G$ be a directed acyclic graph composed by the random variables $X_{1},...,X_{n}$, let $Pa(X_{i})$ be the set of parents of $X_{i}$ in $G$. Then it is said that a distribution $P$ \textit{factorizes} according to $G$ if $P$ can be expressed as a product:
\begin{equation}
    P\left (X_{1},...,X_{n}\right )=\prod_{i=1}^{n}P\left ( X_{i}\mid Pa\left ( X_{i} \right ) \right ) \label{eq:factorBN}
\end{equation}
Equation \ref{eq:factorBN} is derived from the chain rule and the conditional independence relations represented in the Bayesian network. The individual factors, $P\left ( X_{i}\mid Pa\left ( X_{i} \right ) \right )$, are called conditional probabilities distributions, which are commonly estimated from data.

In a Bayesian network, specific evidence may be available; the values for a subset of variables are known, then the posterior probability of the other (unknown) variables can be inferred. This is achieved by propagating the available evidence; several algorithms have been proposed for this purpose, such as probability propagation, variable elimination, junction tree, etc. The interested reader will find the description of those algorithms in \cite{Sucar-2021-PGMbook} Chapter 7.
\\

On the other hand,  CNNs are a type of feed forward neural network that can extract features from data with convolutional structures \citep{Li-2022-ASoCNN}.
CNNs comprise three main types of layers, so a CNN architecture is formed when these layers are stacked. The layers are briefly described next \citep{Li-2022-ASoCNN,Haridas-2019-CNNACS,Khan-2020-ASotRAoDCNN}: \textit{convolutional layers} which are composed by convolutional kernels that work by diving the image into small slices, which helps to extract features' motifs; \textit{pooling layers} which have the purpose of decreasing the complexity of the CNNs, neurons in these layers do not have weights or biases that will be learned in the training process, instead, they perform some fixed function; 
and \textit{fully connected layers}, they are commonly used at the end of the network, which are used for classification or regression. 

\begin{figure}[bt] 
	\centering

    \includegraphics[width=.3\columnwidth]{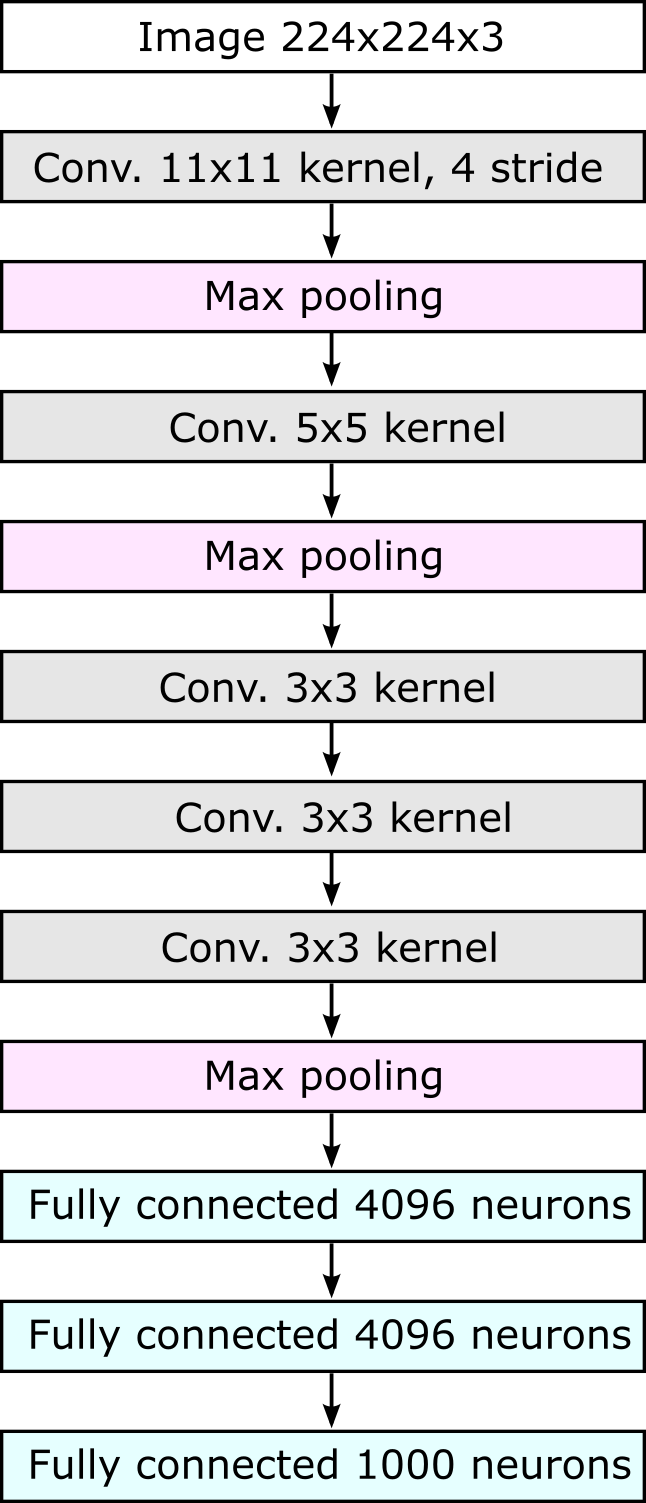}
    \caption{AlexNet architecture\citep{Alex-2012-AlexNet}, which is composed of multiple layers (convolutional, pooling and fully connected).}
  \label{f:alexnet}
\end{figure}

An example of a CNN model is AlexNet \citep{Alex-2012-AlexNet} which was proposed to predict 1000 different classes from the ImageNet LSVRC-2010 contest. AlexNet is composed of five convolutional layers, some of them are followed by pooling layers, and three fully connected layers; AlexNet's architecture is depicted in Fig. \ref{f:alexnet}. 

In this work, pre-trained CNN models are used because they have  several advantages over training one from scratch: (i) pre-trained CNNs have already been trained on large data sets, meaning they have learned feature-rich representations for a wide range of images; (ii) less training data required, this is especially useful when few training data is available for the new task, that is, learning is usually faster and easier than starting from scratch with randomly initialized weights; (iii) less computational resources, because training a network from scratch can be computationally expensive and time consuming, especially if powerful hardware is not available.

\subsection{Evaluation measures}
In this work, three evaluation measures are considered. First, \textit{exact match} \citep{ramirez2016hierarchical,Serrano-Perez}  is the most strict evaluation measure because the prediction must equal the real label set. However, in hierarchical classification, the predictions may be partially correct; in this way, \textit{accuracy} and \textit{hierarchical F-measure} \citep{Silla2011} give a score (greater than zero) if the prediction is partially correct. Let $N$ be the number of instances in the test set, $Y$ be the real subset of classes an instance is associated with, and let $\widehat{Y}$ be the subset of predicted classes. The evaluation measures are described next:
\begin{itemize}
    \item \textbf{Exact Match (EM)}: Percentage of instances classified correctly.
    \begin{equation}
        EM=\frac{1}{N}\sum_{i=1}^{N}1_{Y=\widehat{Y}} \label{e:em}
    \end{equation}
    
    \item \textbf{Hierarchical Accuracy (H. Accuracy)}: Ratio of classes predicted correctly to the union of the real and predicted classes for each instance.
    \begin{equation}
        H. Accuracy=\frac{1}{N}\sum_{i=1}^{N}\frac{\left | Y_{i}\cap \widehat{Y_{i}} \right |}{\left | Y_{i}\cup  \widehat{Y_{i}} \right |}
    \end{equation}
        
    \item \textbf{Hierarchical F-measure (hF)}: F-measure  for hierarchical classification.
    \begin{equation}
        hF=\frac{2*hP*hR}{hP+hR}
    \end{equation}
    \begin{equation*}
        hP=\frac{\sum_{i=1}^{N}\left | Y_{i}\cap \widehat{Y_{i}} \right |}{\sum_{i=1}^{N}\left |  \widehat{Y_{i}} \right |}
    \end{equation*}
    \begin{equation*}
        hR=\frac{\sum_{i=1}^{N}\left | Y_{i}\cap \widehat{Y_{i}} \right |}{\sum_{i=1}^{N}\left |  Y_{i} \right |}
    \end{equation*}
    Where $hP$ and $hR$ are the hierarchical precision and recall, respectively.
\end{itemize}

\subsection{Bayesian and Convolutional Neural Networks} \label{s:bcnn}
\begin{figure}[h!] 
	\centering

    \includegraphics[width=.7\columnwidth]{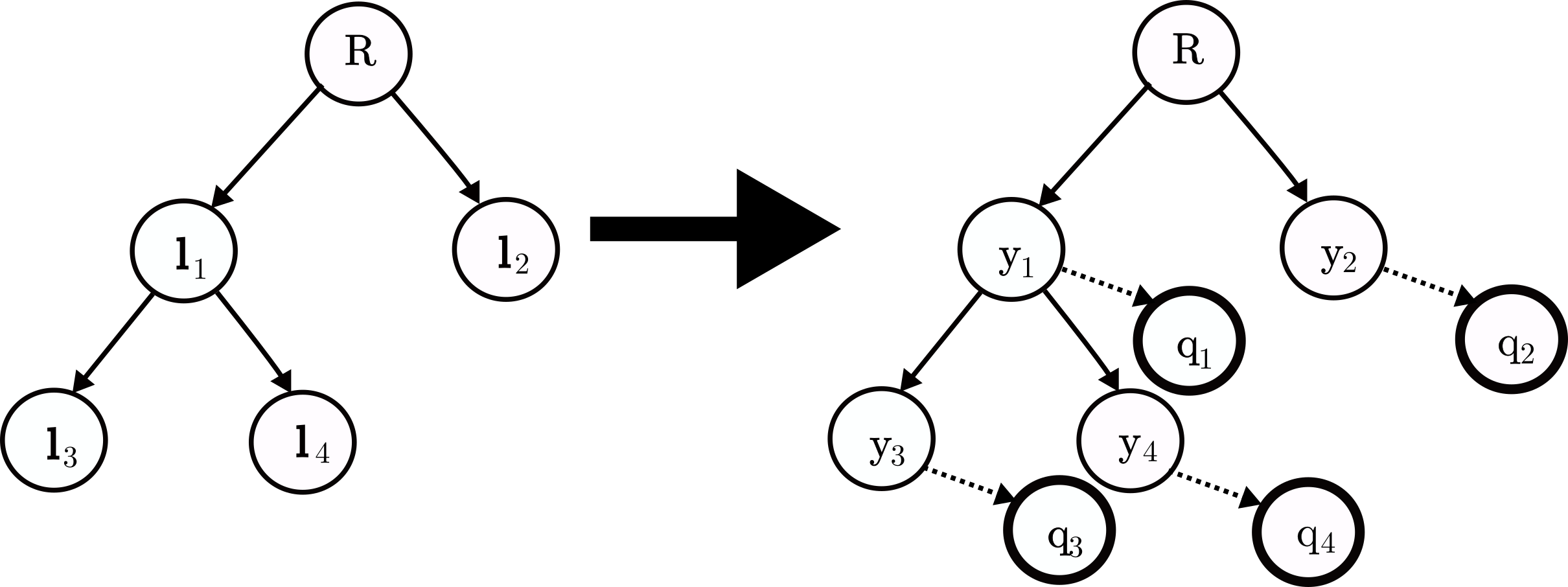}
    \caption{A hierarchy (left) is transformed into a Bayesian network (right).}
  \label{f:hier2bn}
\end{figure}
First of all, the hierarchy is transformed into a Bayesian network \citep{Serrano-Perez}, see Fig. \ref{f:hier2bn}. The Bayesian network comprises two types of random nodes, $y$ and $q$. 
First, $y$ nodes represent the data distribution and maintain the hierarchical constraint in the Bayesian network; hence, for each node of the hierarchy, there is a $y_{i}$ node in the Bayesian network. The parameters for each node $y_{i}$, $P(y_{i}|pa(y_{i}))$, can be estimated by maximum likelihood from the training set as shown in Table \ref{t:cpt_y}, where $pa(y_{i})$ is the parent of $y_{i}$ given by the hierarchy. As it can be seen, if an instance is not associated with $pa(y_{i})$, then it will be no associated to $y_{i}$ with probability one, $P(y_{i}=0|pa(y_{i})=0)=1$, this is what maintains the hierarchical constraint in the Bayesian network.

\begin{table}[h!]
\centering
\caption{Conditional probability table of $P(y_{i}|pa(y_{i}))$. $a$ is the number of instances associated to both $y_{i}$ and its parent, $pa(y_{i})$; $b$ is the number of instances no associated to $y_{i}$ but associated to its parent, $pa(y_{i})$. Laplace smoothing is applied. }
\label{t:cpt_y}
\begin{tabular}{cccc}
                         &                        & \multicolumn{2}{c}{$pa(y_{i})$}                 \\
                         &                        & 1                      & 0                      \\ \cline{3-4} 
\multirow{2}{*}{$y_{i}$} & \multicolumn{1}{c|}{1} & \multicolumn{1}{c|}{$\frac{a+1}{a+b+2}$} & \multicolumn{1}{c|}{0} \\ \cline{3-4} 
                         & \multicolumn{1}{c|}{0} & \multicolumn{1}{c|}{$\frac{b+1}{a+b+2}$} & \multicolumn{1}{c|}{1} \\ \cline{3-4} 
\end{tabular}
\end{table}

Second, each $y_{i}$ node has a child node, $q_{i}$, in the Bayesian network, except the root node. $q$ nodes represent the base classifier output distribution to be expected on instances that were not used as training; that is, $q$ nodes model the behavior of the base classifier for predicting correctly and incorrectly instances. Furthermore, $q$ nodes will receive the predictions of the base classifier, which will be propagated in the Bayesian network. 

Taking into account that the base classifier predicts the probability of being associated with the $i$-th node, the distribution for each node $q_{i}$, $P(q_{i}|y_{i})$, can be modeled parametrically with Gaussian distributions \citep{Barutuoglu-2008-BayesianAHC}. That is, $P(q_{i} \in  \mathbb{R} | y_{i}=0) \simeq N(\mu_{0},\sigma_{0}^{2})$ where $\mu_{0}$ is the mean and $\sigma_{0}^{2}$ is the variance of the predictions of the base classifier in the instances no associated to $y_{i}$ in the validation set; and the same for $P(q_{i} \in  \mathbb{R} | y_{i}=1) \simeq N(\mu_{1},\sigma_{1}^{2})$ but considering the instances associated to $y_{i}$ in the validation set.

\begin{figure}[b!]
	\centering
    \includegraphics[width=.9\columnwidth]{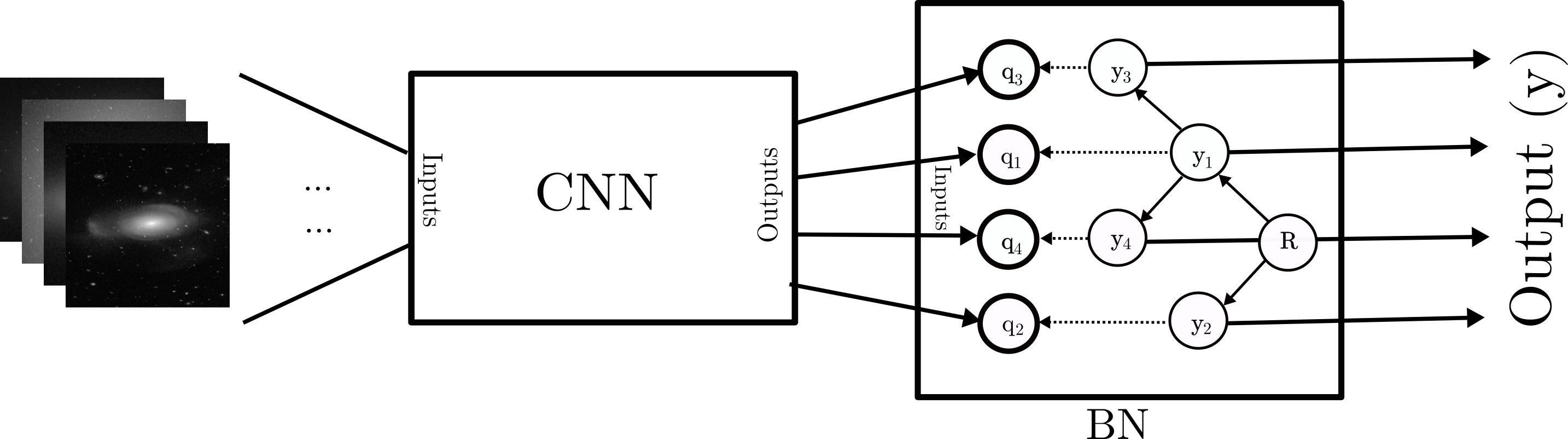}
    \caption{Model of the proposed classifier. It comprises two main modules: a CNN that feeds a Bayesian network. The CNN classifier is fed directly with the images, which outputs the probability for each class (in this example, four classes, $q_{1..4}$). These probabilities are sent to the Bayesian network that optimizes the classification via probabilistic inference. In this simple example, the hierarchy consists of 4 classes, where $y_1$ and $y_2$ are sub-classes of the root, $R$; and $y_3$ and $y_4$ are sub-classes of $y_1$.}
  \label{f:bcnn}
\end{figure}

In this work, the base classifier is a \textit{pretrained} convolutional neural network. CNNs have the advantage that can be trained with \textit{raw} images because they can automate the feature extraction process from the images  \citep{Li-2022-ASoCNN,Altenberger-2018-ANonTSoDCNN}. The CNN in the proposed model  is joined with a dense layer (with \textit{sigmoid} activation function), which has one output for each node of the hierarchy (except the root node) and serves as a classifier. In this way, the training can be carried out only in this last layer; however, the whole CNN can be retrained, too. 
Later, the predictions of the CNN are used to feed the Bayesian network through the $q$ nodes, as shown in Fig. \ref{f:bcnn}.

Finally, the prediction for new instances is obtained from the $y$ nodes and the posterior probabilities. Finally, the top-down (TD)  procedure is used to get only one path of labels. TD starts at the root node and advances toward the child node with the most significant probability, and so on until a leaf node is reached; so this path is returned as the prediction of the model.


To sum up the proposed classifier:
\begin{itemize}
    \item \textit{Training phase}:  The CNN is trained with the labeled images. Then, the parameters of the Bayesian network are estimated. $P(y_{i}|pa(y_{i}))$ can be calculated from the training set, while $P(q_{i}|y_{i})$ can be estimated from the predictions of the CNN in a validation set.

    \item \textit{Prediction phase}: The images are fed to the CNN, the CNN's predictions are sent to the Bayesian network, and the evidence is propagated. Finally, the TD procedure is applied to the posterior probabilities of the $y$ nodes to get the image's final prediction (set of labels).
\end{itemize}


\section{Experiments and Results} \label{s:EandR}

The experiments evaluate the impact of several improvements over the basic CNN classifier, including (i) the incorporation of the hierarchical constraint with a Bayesian network, (ii) selection of the path of labels, (iii) data augmentation and (iv) fine-tuning of the CNN.

In the following experiments, gradient descent is used to train the CNN as the optimizer with a learning rate=0.005 and momentum=0.9. 
All the experiments were executed on a computer with CPU Threadripper 3970X, GPU RTX3090, and RAM 128GB.

\subsection{Datasets and Preprocessing}
\begin{figure}[h!]
	\centering

    \includegraphics[width=.8\columnwidth]{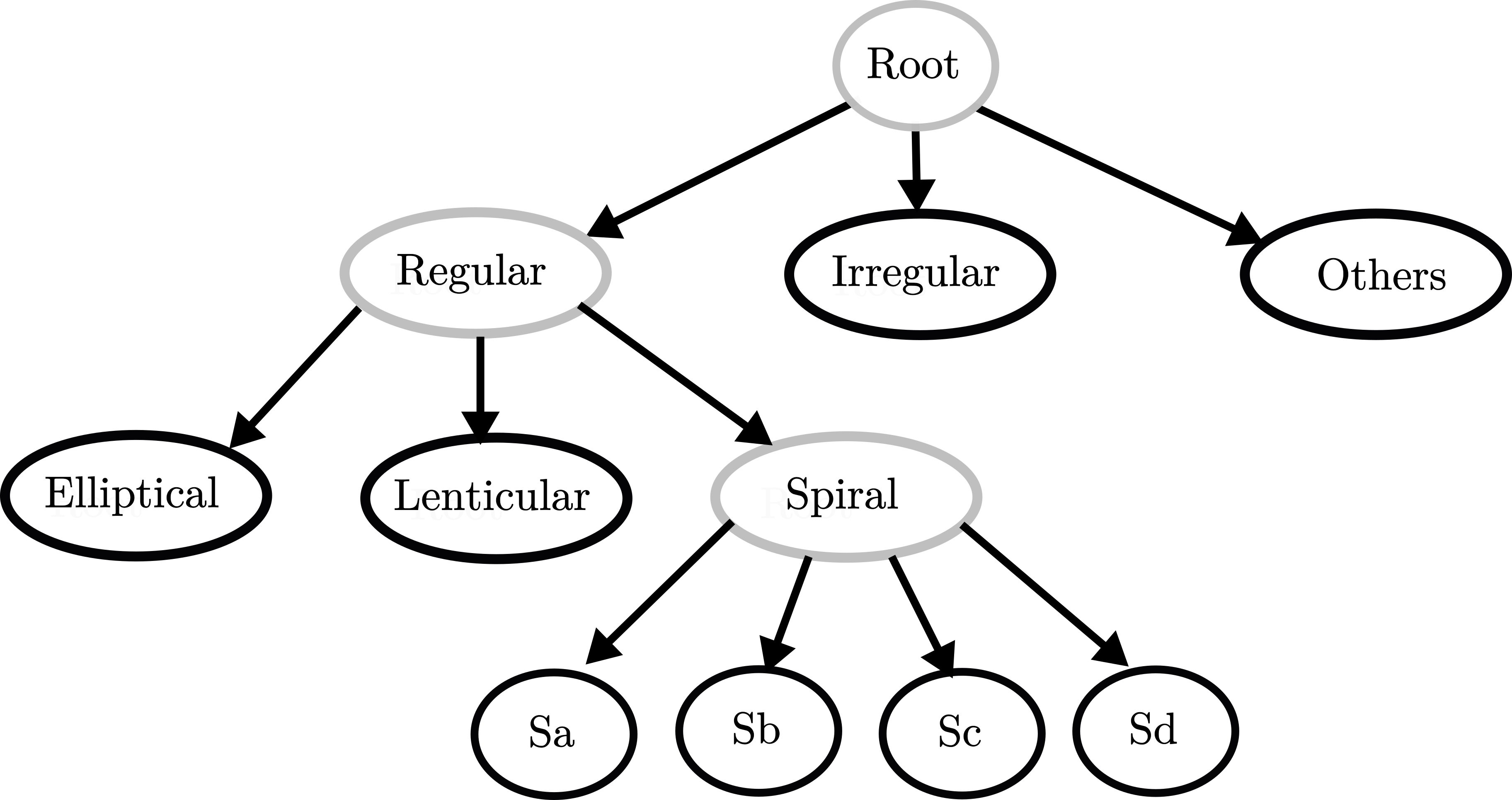}
    \caption{Galaxy hierarchy considered in the experiments.}
  \label{f:h_galaxies}
\end{figure}
We employed a collection of galaxies\footnote{The dataset is available at \url{https://www.kaggle.com/datasets/jonsperez/galaxies-dataset-for-hierarchical-classification}} sourced from the Principal Galaxies Catalog (PGC) and the APM Equatorial Catalogue of Galaxies. Both compilations provide details regarding the morphological and numerical characteristics of each galaxy. Furthermore, we cross-verified the details of each galaxy against the most recent citations available for each of them.   All the images were standardized to size $300 \times 300$ pixels and monochrome. The classes of galaxies are arranged in the hierarchy shown in Fig. \ref{f:h_galaxies}. As it can be seen, the different classes and the hierarchy are based on the Hubble sequence \citep{Hubble-1926-ExtraGN,Hubble-1927-TCoSN}, that is, galaxies of classes \textit{Sa, Sb, Sc} and \textit{Sd} are grouped as \textit{Spiral} galaxies; \textit{Elliptical, Lenticular} and \textit{Spiral} classes are grouped as \textit{Regular} galaxies; and galaxies of classes \textit{Regular, Irregular} and \textit{Others} are grouped in the \textit{root} node. Even though the Hubble sequence does not include the class \textit{Others}, in this dataset the class \textit{Others} contains astronomical objects such as nebulae and star clusters, which are objects that can be found in the dataset. Each image is associated with a single full path of labels (classes); the path starts in the root node and finishes in a leaf node. Table \ref{t:ds_galaxies} presents the number of galaxies per class. 

\begin{table}[tbh]
\caption{Classes of galaxies and the number of images in training, test, and validation sets. *: Images associated with multiple classes are counted only once.}
\label{t:ds_galaxies}
\centering
\begin{tabular}{lcccc}
\hline
\textbf{Class} & \textbf{Total} & \textbf{Training} & \textbf{Test} & \textbf{Validation} \\ \hline
Sa             & 277            & 177               & 56            & 44                  \\
Sb             & 281            & 180               & 56            & 45                  \\
Sc             & 222            & 142               & 45            & 35                  \\
Sd             & 62             & 40                & 12            & 10                  \\
Spiral         & 842            & 539               & 169           & 134                 \\
Elliptical     & 471            & 301               & 94            & 76                  \\
Lenticular     & 387            & 247               & 78            & 62                  \\
Regular        & 1700           & 1087              & 341           & 272                 \\
Irregular      & 181            & 116               & 36            & 29                  \\
Others         & 50             & 32                & 10            & 8                   \\
               &                &                   &               &                     \\
Total*:         & 1931           & 1235              & 387           & 309                 \\ \hline
\end{tabular}
\end{table}

\subsection{Base classifier}
As stated in the section \ref{s:bcnn}, the base classifier is a pre-trained CNN; there are several CNNs freely available; some of the most \textit{popular} are shown in Table \ref{t:cnn_imagenet}. The accuracy in the ImageNet\footnote{ImageNet is an image database organized according to the WordNet hierarchy - https://www.image-net.org/} task is shown for all of them; as it can be seen, the best is EfficientNet V2. At the same time, the well-known Inception v3 is 10 points lower than the first. EfficientNet v2 is selected as the base classifier for the following experiments.

\begin{table}[tbh]
\caption{Some \textit{popular} convolutional neural networks. The top 1 accuracy (percentage) in the ImageNet task and the number of parameters of each one are shown.}
\label{t:cnn_imagenet}
\centering
\begin{tabular}{@{}lcc@{}}
\toprule
\textbf{Model}        & \textbf{Top 1 Accuracy} & \textbf{Parameters} \\ \midrule
EfficientNetV2-xl-21k \citep{Tan-2021-EfficientNetV2} & 87.2                    & 207,615,832          \\
BiT-M R152x4 \citep{Kolesnikov-2020-bit}          & 85.3                    & 928,340,224          \\
Inception ResNet V2 \citep{Szegedy-2017-inception-resnet}   & 80.1                    & 54,336,736           \\
Inception v3 \citep{Palacio-2018-inceptionv3}          & 77.12                   & 21,802,784           \\ \bottomrule
\end{tabular}
\end{table}

    
First, the CNN classifier was trained for 40 epochs over the galaxies training set (Table \ref{t:ds_galaxies}); the evolution of the losses, for training and validation sets, tend to decrease as expected. Nevertheless, the loss for validation set is reduced up to 0.35 on the first 20 epochs, then it is barely reduced. Therefore, to avoid overfitting of the classifier on the training data, the number of epochs was set to 30 in all the next experiments.

\subsection{Incorporation of the Bayesian network}

In this section, the experiment shows whether the inclusion of the Bayesian network helps to improve the classifier's performance. That is, the CNN will be trained without the Bayesian network while, on the other side, the CNN will be trained with the Bayesian network (BCNN); in order to obtain consistent paths, the TD procedure is applied to the output of both classifiers. Table \ref{t:exp_bn} shows the results of this experiment; as it can be seen, making use of the Bayesian network helps to improve the performance of the CNN from 57\% to 64\% in exact match and 77\% to 81\% in hF.

\begin{table}[tbh]
\caption{Results (percentage) of the classifier with and without the Bayesian network, BCNN and CNN, respectively.}
\label{t:exp_bn}
\centering
\begin{tabular}{@{}lccccc@{}}
\toprule
\textbf{Classifier} & \textbf{EM} & \textbf{H. Accuracy} & \textbf{hR} & \textbf{hP} & \textbf{hF} \\ \midrule
CNN                 & 57.62       & 71.45                & 80.38            & 75.03               & 77.61       \\
BCNN              & 64.34       & 76.21                & 81.94            & 80.15               & 81.04       \\
\bottomrule
\end{tabular}
\end{table}

\subsection{Selection of the path of labels}

As described in section \ref{s:bcnn}, the TD procedure is used to obtain the classifier's predictions, that is, the path of labels (classes) to which an instance is associated. However, there are other ways to obtain paths of labels from the probabilities of the nodes, such as \textit{sum of probabilities} (SP) \citep{hernandez2013hybrid} that consists of averaging the probabilities of the nodes that form a path (the path starts on the root node and finishes on a leaf node) and returns the path with the highest score; and \textit{score gain-loose balance} (scoreGLB) \citep{ramirez2016hierarchical} which follows a similar idea than SP but gives weight to each node with respect to its level in the hierarchy, so that higher nodes in the hierarchy have greater weight than lower ones. 

\begin{table}[h!]
\caption{Results (percentage) of the BCNN classifier with different procedures to obtain the path of labels (in bold the best). }
\label{t:exp_paths}
\centering
\begin{tabular}{@{}lccccc@{}}
\toprule
\textbf{Procedure} & \textbf{EM} & \textbf{H. Accuracy} & \textbf{hR} & \textbf{hP} & \textbf{hF} \\ \midrule
TD                 & \textbf{64.34}       & \textbf{76.21}                & \textbf{81.94}            & \textbf{80.15}               & \textbf{81.04}       \\
SP \citep{hernandez2013hybrid}                 & 62.53       & 74.81                & 79.71            & 79.8                & 79.75       \\
scoreGLB \citep{ramirez2016hierarchical}           & 63.31       & 75.3                 & 80.04            & 80.13               & 80.09       \\ \bottomrule
\end{tabular}
\end{table}

Hence, the experiment in this section shows which procedure is the best to obtain the path of labels in the proposed classifier. Table \ref{t:exp_paths} shows the results of the BCNN classifier with different procedures to obtain the predictions of the instances; as seen, the TD procedure got the best performance in all the evaluation measures. We attribute this situation to the fact that SP and scoreGLB are based on the idea that nodes may have a higher probability than their parents/ancestors, which would be beneficial when scoring the paths; however, this case does not occur in the BCNN classifier, because the Bayesian network always keeps the probabilities of the nodes with lower probabilities than their parents. Thus, the TD procedure is the best choice in the proposed classifier.

\subsection{Image augmentation}

The galaxy dataset is unbalanced; that is, some classes have more images associated than others; for instance, classes \textit{Sd} and \textit{others} have 40 and 32 images associated, respectively, while the \textit{elliptical} class has associated 301 images. This makes a difference up to $\sim 10$ times, considering only the training set and the leaf nodes. 

Hence, the following experiment generates artificial images for the classes with few images associated to improve the classifier's general performance. Details of how images are generated are shown in Appendix \ref{a:imgGen}. Table \ref{t:exp_imgAug} shows the results of the BCNN classifier trained with and without image augmentation; as it can be seen, the classifier does not improve its performance when using artificial images as extra data. 
However, when we combine fine-tuning with image augmentation, all evaluation measures improve, as shown in section \ref{s:tl}.

\begin{table}[tbh]
\caption{Results (percentage) of BCNN classifier with and without image augmentation.}
\label{t:exp_imgAug}
\centering
\begin{tabular}{lccccc}
\hline
\textbf{Img. Aug.} & \textbf{EM} & \textbf{H. Accuracy} & \textbf{hR} & \textbf{hP} & \textbf{hF} \\ \hline
No                        & 64.34       & 76.21                & 81.94            & 80.15               & 81.04       \\
Yes                       & 63.82       & 75.95                & 81.72            & 79.93               & 80.82       \\ \hline
\end{tabular}
\end{table}

\subsection{Fine-Tuning} \label{s:tl}
Up to this point, the training of the CNN consisted only of training its last layer, a dense layer with one output for each node of the hierarchy. Therefore, the rest of CNN's layers were only used as feature extractors from the images; but if all its layers are \textit{retrained} (fine-tuned) with the available images, will the classifier's performance improve?

In this experiment, all the layers of the CNN are retrained. Table \ref{t:exp_tl} shows the results of the BCNN classifier allowing or not retraining all the layers of the CNN. As can be seen, retraining the whole CNN helps to improve the classifier's performance in all the evaluation measures. Additionally, an experiment with both retraining all the CNN's layers and with image augmentation is carried out; the results are shown in the last row (Yes+ImgA); as can be seen, the combination of both gives the best results.

\begin{table}[tbh]
\caption{Results (percentage) of BCNN classifier retraining or not all the CNN's layers. ImgA: plus image augmentation. (In bold the best.) }
\label{t:exp_tl}
\centering
\begin{tabular}{lccccc}
\hline
\textbf{TL} & \textbf{EM}    & \textbf{H. Accuracy} & \textbf{hR} & \textbf{hP} & \textbf{hF}    \\ \hline
No          & 64.34          & 76.21                & 81.94            & 80.15               & 81.04          \\
Yes         & 65.89          & 77.43                & 82.16            & 82.25               & 82.21          \\
Yes+ImgA    & \textbf{67.18} & \textbf{78.68}       & \textbf{84.17}   & \textbf{82.33}      & \textbf{83.24} \\ \hline
\end{tabular}
\end{table}

\subsection{Comparison of BCNN against CNN models} \label{s:comparison}
This section compares the proposed classifier against Deep Galaxy 2 \citep{Khalifa-2018-DeepGalaxyV2} and different CNN models. In the same way as BCNN,  all of them were joined with a dense layer with one output per class and \textit{sigmoid} activation functions; also, the models were optimized with the \textit{Gradient descent (with momentum)} optimizer and \textit{binary cross-entropy} as the loss. Additionally, in order to obtain consistent predictions from all CNN models, the TD procedure is applied to their predictions.

Table \ref{t:comparison} shows the results of the proposed classifier, BCNN, and the different models. BCNN outperforms the rest of the classifiers in all the evaluation measures. Furthermore, when BCNN is trained with image augmentation and fine-tuning the performance  increased from 64\% to 67\% in exact match and 81\% to 83\% in hF.

\begin{table}[tbh]
\caption{Results (percentage) of the BCNN classifier compared with several CNN models. ImgA: plus image augmentation; FN: plus fine-tuning. (In bold the best.)}
\label{t:comparison}
\centering
\begin{tabular}{lccccc}
\hline
\textbf{Classifier}    & \textbf{EM}    & \textbf{H. Accuracy} & \textbf{hR} & \textbf{hP} & \textbf{hF}    \\ \hline
Deep Galaxy v2      & 16.54                           & 41.34                & 63.99            & 49.61               & 55.89       \\
Inception v3           & 48.32          & 64.84                & 76.14            & 69.27               & 72.54          \\
Inception ResNet v2    & 42.12          & 60.1                 & 73.91            & 64.62               & 68.95          \\
BiT m-r152x4           & 52.97          & 68.07                & 72.69            & 75.29               & 73.96          \\
EfficientNet V2-xl-21k & 57.62          & 71.45                & 80.38            & 75.03               & 77.61          \\ \hline
BCNN                   & 64.34          & 76.21                & 81.94            & 80.15               & 81.04          \\
BCNN(ImgA,FN)          & \textbf{67.18} & \textbf{78.68}       & \textbf{84.17}   & \textbf{82.33}      & \textbf{83.24} \\ \hline
\end{tabular}
\end{table}

\subsection{Discussion}

From the experiments, we can conclude that all the elements --Bayesian network hierarchical classifier, data augmentation, and fine-tuning-- contribute to improve the performance of the basic CNN classifier in all measures. There is an approx. $10$ point improvement on exact match and $7$ points for the hierarchical F-measure over the baseline. Of the different elements, the Bayesian network provides the most significant impact on performance, $\sim 7$ points (from 57\% to 64\%) in exact match and $\sim 3$ points (77\% to 81\%) in hF.

\section{Conclusions and Future Work}\label{s:CandFW}

Morphological galaxy classification is still a challenging problem. In this work, we proposed a novel approach, Bayesian and Convolutional Neural Networks, for morphological galaxy classification. This method combines a convolutional neural network trained with different classes of galaxies and a Bayesian network that represents the hierarchy of each type analyzed by CNN. The BN helps to improve the performance by enforcing the hierarchical constraint via probabilistic inference. Further improvements are obtained by image augmentation and fine-tuning the CNN. 

The proposed approach was evaluated on images collected from different repositories considering ten classes of galaxies.
The experiments show that all the BCNN components help the model improve its performance. Furthermore, BCNN got the best performance when compared to other CNN models.

In future work, a semi-supervised approach could be carried out to incorporate unlabeled images of galaxies.

\backmatter

\bmhead{Supplementary information}
Not applicable.

\bmhead{Acknowledgments}

J. Serrano-Pérez acknowledges the support from CONAHCYT scholarship number (CVU) 84075.

\section*{Declarations}

\begin{itemize}
\item Funding: J. Serrano-Pérez was supported by CONAHCYT scholarship number (CVU) 84075.
\item Conflict of interest/Competing interests: The authors declare that they have no known competing financial interest, conflict of interest, or otherwise.
\item Ethics approval: Not applicable.
\item Consent to participate: Not applicable.
\item Consent for publication: Not applicable.
\item Availability of data and materials: The dataset is made available to the scientific community: \url{https://www.kaggle.com/datasets/jonsperez/galaxies-dataset-for-hierarchical-classification} 
\item Code availability: The code is available at: \url{https://github.com/jona2510/BCNN}

\item Authors' contributions (CRediT): \textbf{Jonathan Serrano-Pérez}: Conceptualization, Software, Validation, Formal analysis, Writing-Original Draft, Writing-Review \& Editing. \textbf{Raquel Díaz Hernández}: Conceptualization, Formal analysis, Writing-Original Draft, Writing-Review \& Editing. \textbf{L. Enrique Sucar}: Conceptualization, Formal analysis, Resources, Writing-Original Draft, Writing-Review \& Editing, Supervision.
\end{itemize}

\begin{appendices}

\section{Image augmentation} \label{a:imgGen}
Data augmentation is carried out to \textit{balance} the leaf nodes. It consists of applying some operations to the available images, such as flip, scale, rotation, etc., to obtain different images. The detailed procedure is described next.

\begin{figure}[tbh]
	\centering

    \includegraphics[width=.6\columnwidth]{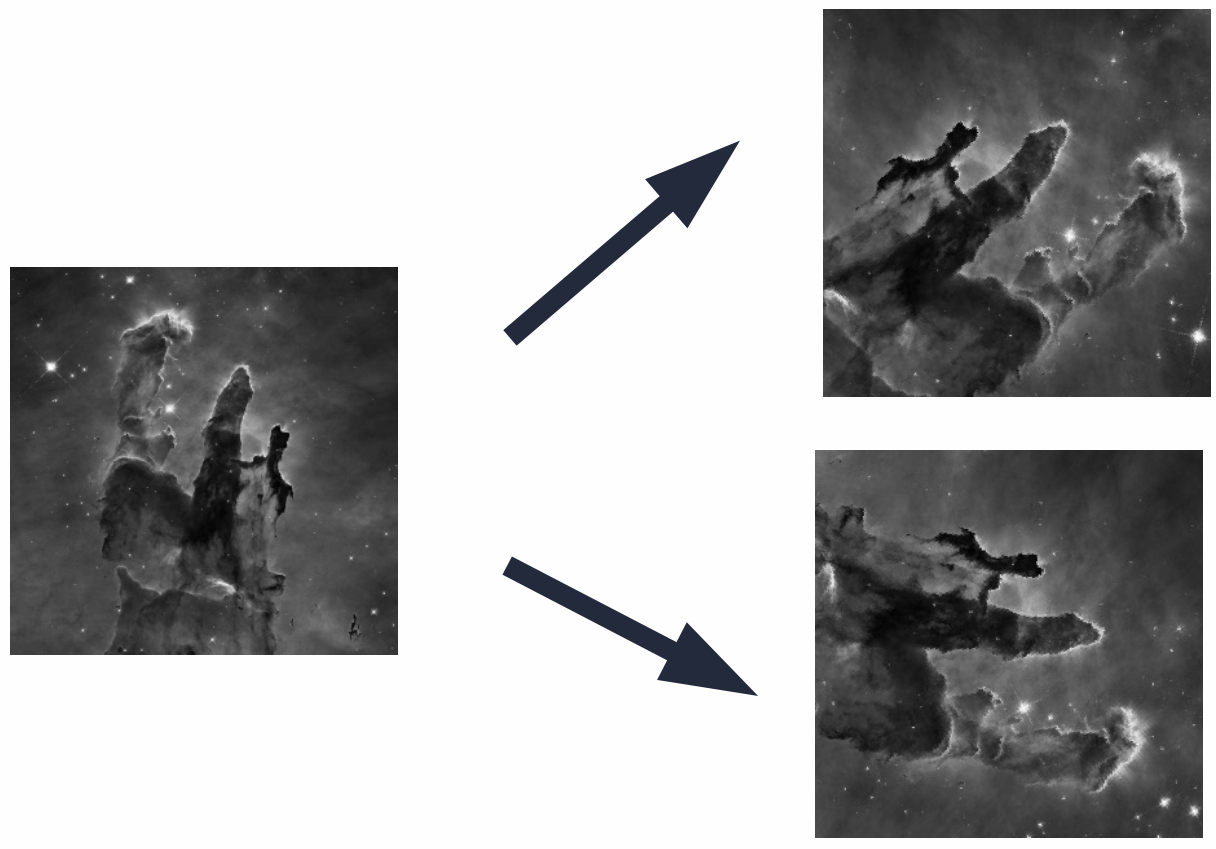}
    \caption{example of generation of artificial images. The image on the left is the available one; the two on the left are generated images after applying some operations to the image on the left. }
  \label{f:imgAug_ex}
\end{figure}

First, let $LMAX$ be the number of associated instances to the leaf node with the most significant number of related instances.
Then, for each $l$ leaf node with  $m_{l}$ instances associated: generate $n$ images so that $n+m_{l} = LMAX*0.95$; the $n$ generated images will be associated to $l$ and all its ancestors given by the hierarchy. Image generation follows the next pipeline (iterating over the images associated with the $l$ node):

\begin{enumerate}
    \item Flip (either horizontal or vertical) with a probability of 0.5.

    \item  Shift both x and y axis in the range [-10, 10] (percent) with probability one.

    \item Scale in the range [-30, 30] (percent) with probability one.

    \item Rotate in the range [0, 360] (degrees) with probability one.

    \item To fill the \textit{voids} that may appear in the images due to the different transformation, the strategy \textit{border reflect}\footnote{BORDER\_REFLECT\_101 from opencv.org} is used to fill those pixels.

    \item  Resize to 300x300 pixels.
\end{enumerate}
An example is shown in Fig. \ref{f:imgAug_ex}. Two images are generated from the image on the left.

\end{appendices}


\bibliography{sn-bibliography}

\end{document}